# A model-based approach to assess the effectiveness of pest biocontrol by natural enemies

Mamadou Ciss[a], Sylvain Poggi[b], Mohamed-Mahmoud Memmah[a], Pierre Franck[a], Marie Gosme[c], Nicolas Parisey[b], Lionel Roques[d]

[a]INRA, UR 1115 Plantes et Systèmes de culture Horticoles, F-84000 Avignon, France
[b]INRA UMR 1349 IGEPP, F-35653 Le Rheu Cedex, France
[c]INRA, UMR 1230 SYSTEM, F-34060 Montpellier Cedex 2, France
[d]INRA, UR 546 Biostatiques et Processus Spatiaux, F-84000 Avignon, France

## Abstract

*Main goal:* The aim of this note is to propose a modeling approach for assessing the effectiveness of pest biocontrol by natural enemies in diversified agricultural landscapes including several pesticide-based management strategies. Our approach combines a stochastic landscape model with a spatially-explicit model of population dynamics. It enables us to analyze the effect of the landscape composition (proportion of semi-natural habitat, non-treated crops, slightly treated crops and conventionally treated crops) on the effectiveness of pest biocontrol. Effectiveness is measured through environmental and agronomical descriptors, measuring respectively the impact of the pesticides on the environment and the average agronomic productivity of the whole landscape taking into account losses caused by pests.

*Conclusions:* The effectiveness of the pesticide, the intensity of the treatment and the pest intrinsic growth rate are found to be the main drivers of landscape productivity. The loss in productivity due to a reduced use of pesticide can be partly compensated by biocontrol. However, the model suggests that it is not possible to maintain a constant level of productivity while reducing the use of pesticides, even with highly efficient natural enemies. Fragmentation of the semi-natural habitats and increased crop rotation tend to slightly enhance the effectiveness of biocontrol but have a marginal effect compared to the predation rate by natural enemies.

# Organization of the note

Section I is devoted to
> (1) the definition of the landscape models and the main types of landscape compositions that are compared in this note. The models presented in this section allow us to generate dynamic stochastic landscapes made of several types of land-uses with a control of the proportion occupied by each type of land-use, of the landscape fragmentation, and of the temporal changes in land-use;
> (2) the definition of a spatially-explicit dynamical system describing pest and natural enemy interactions.

Section II is devoted to the definition of environmental and agronomical performance criteria.
Section III describes:
> (1) the range of parameter values which are used in our numerical computations;
> (2) the design of the numerical experiments.

The results are presented in Section IV.

# I. The models

## I.1 Stochastic landscape models

*General framework.* The landscape model used in this study is inspired from statistical physics and corresponds to an extension of the stochastic model proposed by Roques and Stoica (2007). It is a neutral landscape model in the sense that the value associated to each position in the landscape (the land-use in the present case) is a random variable (Gardner, 1987). The initial model proposed in Roques and Stoica (2007) only generated static and binary habitat/matrix landscapes. The model presented here allows generating dynamic stochastic landscapes made of several types of land-uses with a control of the proportion occupied by each type of land-use, of the landscape fragmentation, and of the temporal changes in land-use.

*State space.* Landscapes are defined on a lattice Z (grid with cells), each cell being attached to a given land-use. We assume that the lattice is of size $L \times L$, so that it corresponds to the set of all couples $x = (i, j) \epsilon [1, L] \times [1, L]$. A landscape $\omega$ is a random field defined on the lattice Z, and that assigns each site $z$ a value $k = 1,2,3,4$ corresponding to the land-use. The proportion of each land-use being fixed (respectively denoted by $p_1, p_2, p_3, p_4$), the state space $\Omega$ is defined as the set of all of the possible landscapes $\omega$ corresponding to these fixed proportions.

*Static landscape model (MULTILAND).* In a first step, we define a static landscape model with 4 types of land-uses. The landscape model is based on the Gibbs measure $P$ describing the probability associated to each landscape $\omega$:

$$P(\omega) = \frac{1}{Q} e^{-\beta s(\omega)}$$

where $Q$ is a normalization constant and $s(\omega)$ is related to the level of fragmentation of the first type of land-use in the landscape $\omega$. More precisely, $s(\omega)$ is the number of pairs of similar neighbors in the region occupied by the land-use 1 (see Roques, 2015 for a mathematical definition of $s(\omega)$). Thus, a landscape pattern $\omega$ is all the more aggregated as $s(\omega)$ is high, and all the more fragmented as $s(\omega)$ is small. Based on the quantity $s(\omega)$, and using a method introduced in Garnier et al. (2012) we compute a fragmentation index $fr(\omega)$ in [0,1] ($fr(\omega)$=1 corresponds to the most fragmented landscapes, see Roques, 2015). The parameter $\beta$ allows us the control of the fragmentation level of the landscapes drawn from the Gibbs measure $P$: the landscape model tends to produce more aggregated landscapes as $\beta > 0$ increases and more fragmented landscapes as $\beta < 0$ decreases. In our study, samples from the Gibbs measure $P$ were obtained using a Metropolis-Hastings algorithm (Robert and Casella, 2004), with adaptive values of $\beta$ to reach a priori fixed fragmentation indices. A Matlab® source code of the software MULTILAND is available at http://multiland.biosp.org.

*Land-use distributions.* We consider 24 x 24 lattice landscapes composed of four land-uses: (i) land-use 1 refers to semi-natural habitats, (ii) land-use 2 refers to crops with no pesticide, (iii) land-use 3 refers to slightly treated crops (moderate use of pesticides), and (iv) land-use 4 refers to conventionally treated crops.

Then, we consider five types of land-use distributions $\Omega_B$ (balanced), $\Omega_{SN}$ (half occupied by semi-natural habitats), $\Omega_{NT}$ (half occupied by not treated crops), $\Omega_{ST}$ (half occupied by slightly treated crops) and $\Omega_{CT}$ (half occupied by conventionally treated crops), defined by the proportions $p_1, p_2, p_3, p_4$ associated with each land-use, see Table 1. Figure 1 shows examples of landscapes generated with our model.

|  | $p_1$ | $p_2$ | $p_3$ | $p_4$ |
|---|---|---|---|---|
| $\Omega_B$ | 1/4 | 1/4 | 1/4 | 1/4 |
| $\Omega_{SN}$ | 1/2 | 1/6 | 1/6 | 1/6 |
| $\Omega_{NT}$ | 1/6 | 1/2 | 1/6 | 1/6 |
| $\Omega_{ST}$ | 1/6 | 1/6 | 1/2 | 1/6 |
| $\Omega_{CT}$ | 1/6 | 1/6 | 1/6 | 1/2 |

**Table 1:** Definition of the five types of land-use distributions $\Omega_B$ (balanced), $\Omega_{SN}$ (half occupied by semi-natural habitats), $\Omega_{NT}$ (half occupied by not treated crops), $\Omega_{ST}$ (half occupied by slightly treated crops) and $\Omega_{CT}$ (half occupied by conventionally treated crops). The $p_i$ values associated with each generated landscape correspond to the proportions of semi-natural habitats ($p_1$), crops with no pesticide ($p_2$), slightly treated crops ($p_3$), and conventionally treated crops ($p_4$) respectively.

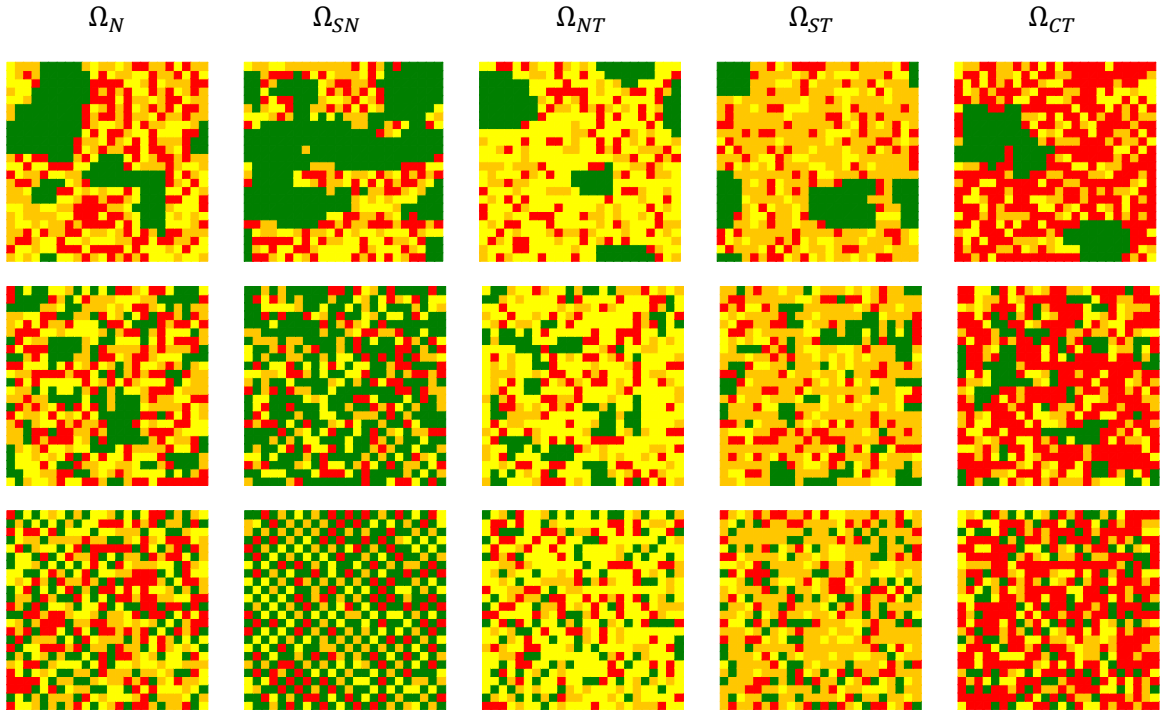

**Figure 1:** Landscapes generated with our algorithm. From left to right: balanced distribution $\Omega_N$, half occupied by semi-natural $\Omega_{SN}$, half occupied by non-treated $\Omega_{NT}$, half occupied by slightly treated $\Omega_{ST}$ and half occupied by conventionally treated $\Omega_{CT}$. Top row: fragmentation index $fr = 0.1$; middle row: fragmentation index $fr = 0.5$; bottom row: fragmentation index $fr = 0.9$. The colors are: green for semi-natural habitats; yellow for non-treated crops; dark yellow for slightly treated crops; and red for conventionally treated crops.

*Dynamic landscape model.* Our algorithm builds sequences of landscapes $(\omega_n)$ associated with a sequence of times $(t_n)$ corresponding to changes in the land-uses. We assume that changes in landscapes are subject to the following constraints: (i) the time increments are supposed to be constant and equal to one year ($t_{n+1} = t_n + 1$); (ii) at each change ($n \to n + 1$), a proportion $\mu$ of the cultivated landscape is modified (this corresponds to an expected period of $1/\mu$ years between two changes for each crop in the lattice); (iii) the locations of semi-natural habitats (land-use of type 1) remain unchanged; (iv) the proportions associated with each type of land-use remain unchanged. To generate a sequence of landscapes satisfying the above constraints, we first generate an initial landscape $\omega_0$ with the static MULTILAND model, using a Metropolis-Hastings MCMC algorithm. Then, given $\omega_n$, the subsequent landscape $\omega_{n+1}$ is generated by continuing the Metropolis-Hastings MCMC algorithm until a proportion $\mu$ of the landscape is modified.

## I.2. Population dynamics

We use a generalized Lotka-Volterra model to describe the populations of pests and their natural enemies. The approach adopted here falls in the framework of lattice dynamical systems. The main difference between this approach and the more classical reaction-diffusion approach relies in the fact that we work on a discrete space, which may be more suited to the landscapes defined in the previous section.

Using the lattice $Z$ defined in Section I.1, we describe the dynamics of the pest population $P_t(x)$ and of the population of natural enemies $N_t(x)$ by the following equations:

$$\begin{cases} P'_t(x) = d_P \, D[P_t(x)] + f_P(t, x, P_t(x)) - \rho(t,x)P_t(x) - \alpha_1 P_t(x)N_t(x) \\ N'_t(x) = d_N \, D[N_t(x)] + f_N(t, x, N_t(x)) - \rho(t,x)N_t(x) + \alpha_2 P_t(x)N_t(x) \end{cases} \quad (1)$$

where $D[.]$ is the discrete Laplace operator modeling the movements of the individuals and defined as: $D[U(x)] = D[U_{ij}] = (U_{i+1j} + U_{i-1j} + U_{ij+1} + U_{ij-1} - 4U_{ij})/\delta_x^2$. This operator ensures that, during a time interval of length $\delta_t \ll 1$, at each position $x = (i,j)$, a proportion $4d_P\delta_t/\delta_x^2$ of the pest population (resp. $4\,d_N\delta_t/\delta_x^2$ of the natural enemy population) moves to the adjacent cells. Thus $d_P$ and $d_N$ directly control the mobility of the pest and natural enemy populations. Here, $\delta_x = 1/n$ is the length of a unit cell in the landscape. We assume periodic conditions at the boundaries of the lattice.

The terms $f_P(t, x, P_t(x))$ and $f_N(t, x, N_t(x))$ correspond to the pest and natural enemy growth functions, while $\rho(t,x)P_t(x)$ and $\rho(t,x)N_t(x)$ describe the pest and natural enemy death rates caused by the use of pesticides. We assume that harvesting occurs at the end of each year (i.e., for integer values of $t$), and that during the first half of each year, $f_P(t, x, P_t(x)) = 0$ due to the absence of resource and $\rho(t,x)$ (no treatment). The precise forms of these functions, depending on the land-use are detailed in Table 2. The interaction terms $-\alpha_1 P_t(x)N_t(x)$ and $\alpha_2 P_t(x)N_t(x)$ describe the effects of predation on the pest and natural enemy growth rates, respectively.

The system (1) has been scaled so that the carrying capacities of $P$ and $N$ are both equal to 1 (this means that the population densities are expressed in units of their respective carrying capacities). We also assume that $\alpha_1 = \alpha_2 = \alpha$ (see Section III.1). The initial condition at $t = 0$ is $N_0(x) = 1$ in the semi-natural habitats and $N_0(x) = 0$ in the crops and $P_0(x) = 0$ everywhere (no pests). The pests are introduced at the beginning of year 3, with $P_3(x) = 0.1$.

| | $f_P(t,x,P)$ | | $f_N(t,x,N)$ | | $\rho(t,x)$ | |
|---|---|---|---|---|---|---|
| Time span | $\left(\lfloor t \rfloor, \lfloor t \rfloor + \frac{1}{2}\right)$ | $\left[\lfloor t \rfloor + \frac{1}{2}, \lfloor t \rfloor + 1\right]$ | $\left(\lfloor t \rfloor, \lfloor t \rfloor + \frac{1}{2}\right)$ | $\left[\lfloor t \rfloor + \frac{1}{2}, \lfloor t \rfloor + 1\right]$ | $\left(\lfloor t \rfloor, \lfloor t \rfloor + \frac{1}{2}\right)$ | $\left[\lfloor t \rfloor + \frac{1}{2}, \lfloor t \rfloor + 1\right]$ |
| Land use: CT | 0 | $r_P P(1-P)$ | $-N/\gamma$ | $-N/\gamma$ | 0 | $2\nu$ |
| Land use: ST | 0 | $r_P P(1-P)$ | $-N/\gamma$ | $-N/\gamma$ | 0 | $\nu$ |
| Land use: NT | 0 | $r_P P(1-P)$ | $-N/\gamma$ | $-N/\gamma$ | 0 | 0 |
| Land use: SN | 0 | 0 | $r_N N(1-N)$ | $r_N N(1-N)$ | 0 | 0 |

**Table 1:** Values of the growth functions $f_P(t,x,P_t(x))$ and $f_N(t,x,N_t(x))$ and of the pesticide-induced death rate in each land-use type (*SN*: semi-natural habitat, *NT*: non-treated crop, *ST*: slightly treated crop, *CT*: conventionally treated crop) $r_P$ is the pest intrinsic growth rate in the crops in the absence of pesticide, $\gamma$ the natural enemy life expectancy in the absence of resource and $r_N$ the natural enemy birth rate in semi-natural habitats. $\nu$ describes the effect of the pesticide on the pest and the natural enemies. The pesticides effects are assumed (i) to be the same on the pests and the natural enemies and (ii) to be twice as large in conventionally treated crops (land-use type *CT*) than in slightly treated crops (land-use type *ST*) and zero elsewhere. $\lfloor t \rfloor$ is the nearest integer less than or equal to $t$.

## II. Performance criteria

### II.1 Environmental criterion

The environmental criterion is linked to the untreated proportion of the study site. With the assumptions of Section I.2, the quantity of pesticide in slightly treated crops is half the quantity in the conventionally treated crops. Thus, the environmental criterion was expressed as follows:

$$ENVI = \frac{Total\ area\ -\ slightly\ treated\ area/2 - conventionally\ treated\ area}{Total\ area}.$$

This formula could be rewritten as follows:

$$ENVI = 1 - p_3/2 - p_4,$$

where $p_3, p_4$ are respectively the proportions of slightly treated crops and conventionally treated crops. Thus, if no pesticide were used $ENVI = 1$ while if the whole study site was conventionally treated $ENVI = 0$.

## II.2 Agronomical criterion

The agronomical criterion reflects the average productivity of the landscape. We make the assumption that, in the absence of pests, the productivity, measured at the end of each year, is proportional to the cultivated area. In the presence of pests, it is weighted by $1 - P_t(x)$, which corresponds to the unaffected resource (we recall that the pest density is expressed in unit of the carrying capacity, thus $P_t(x)$ cannot reach values larger than 1). With these assumptions, the agronomical criterion is formulated as follows:

$$AGRO = \frac{1}{T} \frac{Unit\ area}{Total\ area} \sum_{t=1}^{T} \sum_{x\ in\ crops} (1 - P_t(x)) dt$$

with $Unit\ area = \delta_x^2$, and $T$ is the upper time limit for our simulations which was fixed to 30 (years). Note that, with these assumptions, the agronomical criterion reaches its maximum value 1 in the absence of pests and if the cultivated area occupies the entire study site. Inversely, the agronomical criterion reaches its minimum value 0 if, for each time $t$, the level of pests reaches its maximum ($P_t(x) = 1$) or if the study site is exclusively made of semi-natural habitats. Note that the environmental and agronomical criteria really play antagonistic roles in the sense that (i) if the whole study site was made of semi-natural regions, $ENVI = 1$ and $AGRO = 0$; (ii) if all of the study site was conventionally treated and if all the pests were eradicated, $ENVI = 0$ and $AGRO = 1$.

## III. Simulation study

### III.1 Parameter values

*Intrinsic growth rates* $(r_P, r_N)$. Consider a Malthusian non-spatial model $P' = r_P P$, corresponding to the absence of intraspecific and interspecific interactions, of dispersion and of pesticides. During 1/2 year (corresponding to the period between the emergence of the pest and the crop harvest), the population is increased by a factor $e^{r_P/2}$. $r_P$ was set to 2ln(2), 2ln(50) and 2ln(100), corresponding to a 2-times, 50-times and hundred-times population increase in each period. We assumed the increase factor for the natural enemy in semi-natural habitats to be 2 in one year (contrary to the pest, the natural enemy can grow even during the first half of each year as semi-natural habitats have permanent resources), meaning that $r_N = \ln(2)$.

*Diffusion coefficients* $(d_P, d_N)$. At each time step $\delta_t \ll 1$, a proportion $4\ d\ \delta_t/\delta_x^2$ of the population (pest or natural enemy) situated in a cell $x$ moves into the surrounding cells. We assumed that between 0.1% and 1% of the population moved to the surrounding cells per day, corresponding approximatively to $d_P, d_N \in \frac{1}{n^2}\{0.1, 0.5, 1\}$, with $\delta_t = 1/365$ year = 1 day and $\delta_x = \frac{1}{n}$.

*Effect of pesticide (v).* We assumed that the mortality rate induced by the use of pesticide can reach levels comparable to the pest growth rates: $v \in 2\{\ln(2), \ln(50), \ln(100)\}$.

*Life expectancy of the natural enemies on the crops ($\gamma$).* We assumed a life expectancy of $\gamma = 1/2$ year in the crops, in the absence of pests.

*Interaction terms ($\alpha_1, \alpha_2$).* Consider an isolated and non-treated crop $x$. In the absence of dispersion, and assuming that $f_P(t, x, P) = r_P P(1 - P)$, the equation (1) becomes:

$$\begin{cases} P'_t = r_P P(1 - P) - \alpha_1 P_t N_t, \\ N'_t = -\frac{N_t}{\gamma} + \alpha_2 P_t N_t. \end{cases}$$

The steady states of this system are $(0,0), (1,0)$ and $\left(\frac{1}{\gamma \alpha_2}, \frac{r_P}{\alpha_1}\left(1 - \frac{1}{\alpha_2 \gamma}\right)\right)$, which corresponds to a coexistence state between the pest and its natural enemy. This last state only exists (and is stable) if $\alpha_2 \gamma > 1$. Here, we chose $\alpha_2$ such that the steady state corresponding to the pest population is equal to 25% (high effect of predation) or 75% (moderate effect of predation) of the carrying capacity. Namely, $\alpha_2 \in \frac{1}{\gamma}\left\{\frac{4}{3}, 4\right\}$. We also consider the case where there is no effect of predation: $\alpha_2 = 0$. Note that equilibrium density of the prey in the above reduced system does not depend on $\alpha_1$. For the sake of simplicity, we assume that $\alpha_1 = \alpha_2 = \alpha$.

### III.2 Numerical experiments

Numerical simulations were conducted for the five contrasted distributions of land uses described in Table 1 and Figure 1. Each of these land-use distributions corresponds to a specific value of the environmental criterion: $ENVI(\Omega_{CT}) = 0.420$, $ENVI(\Omega_{ST}) = 0.580$, $ENVI(\Omega_B) = 0.625$, $ENVI(\Omega_{NT}) = 0.750$ and $ENVI(\Omega_{SN}) = 0.750$.

For each land-use distribution, each value of the fragmentation index $fr$ (Table 2), and each value of the crop rotation index µ, we generated 30 sequences of landscapes with the dynamic landscape model presented in Section I.1. This makes a total of $5 * 3 * 3 * 30 = 1350$ sequences of landscapes.

The dynamics of the pest and of its natural enemy were simulated in these landscapes for three different values of each of the five parameters $r_P$, $v$, $\alpha$, $d_P$ and $d_N$ (Table 2), i.e. $3^5 = 243$ combinations of the biological/pesticide parameters. This corresponded to a total of 1350 *sequences of landscapes* $*$ 243 *parameter combinations* $= 328050$ simulations. These simulations were performed with Matlab®.

| Parameter | Description | Unit | Values |
|---|---|---|---|
| **Dynamic landscape model** | | | |
| $fr$ | Fragmentation index | Dimensionless | {0.1,0.5,0.9} |
| $\mu$ | Crop rotation index | Dimensionless | {0,0.1,0.5} |
| $n$ | Size of the lattice $Z$ | Dimensionless | 24 |
| **Model of population dynamics** | | | |
| $d_P$ | Pest diffusion coefficient | Unit area.year$^{-1}$ | $\frac{1}{n^2}\{0.1,0.5,1\}$ |
| $d_N$ | Natural enemy diffusion coefficient | Unit area.year$^{-1}$ | $\frac{1}{n^2}\{0.1,0.5,1\}$ |
| $r_P$ | Pest intrinsic growth rate | year$^{-1}$ | $2\{\ln(2),\ln(50),\ln(100)\}$ |
| $\alpha$ | Predation index | Unit area indiv$^{-1}$ year$^{-1}$ | $\frac{1}{\gamma}\{0,4/3,4\}$ |
| $v$ | Pesticide effect | year$^{-1}$ | $2\{\ln(2),\ln(50),\ln(100)\}$ |
| $\gamma$ | Life expectancy of the natural enemy | year | 1/2 |

**Table 2:** Parameter values used in our simulations. These values are explained in Section III.1.

# IV. Results

## IV. 1 Effect of land-use distribution on landscape productivity

The agronomical performance criterion *AGRO* tends to be negatively correlated with the environmental criterion *ENVI* across the five land-use distributions: the mean value for AGRO criterion over the 65610 simulations of each land-use distribution is 0.66, 0.63, 0.55, 0.5 and 0.38, for $\boldsymbol{\Omega_{CT}}$, $\boldsymbol{\Omega_{ST}}$, $\boldsymbol{\Omega_B}$, $\boldsymbol{\Omega_{NT}}$ and $\boldsymbol{\Omega_{SN}}$ respectively (Figure 2), while $ENVI(\boldsymbol{\Omega_{CT}}) = 0.420$, $ENVI(\boldsymbol{\Omega_{ST}}) = 0.580$, $ENVI(\boldsymbol{\Omega_B}) = 0.625$, $ENVI(\boldsymbol{\Omega_{NT}}) = 0.750$ and $ENVI(\boldsymbol{\Omega_{SN}}) = 0.750$. However, the variability among the *AGRO* criterion values obtained with different sets of parameters remains high, showing that for a given level of the environmental criterion, there might be room for improvement of the agronomical criterion.

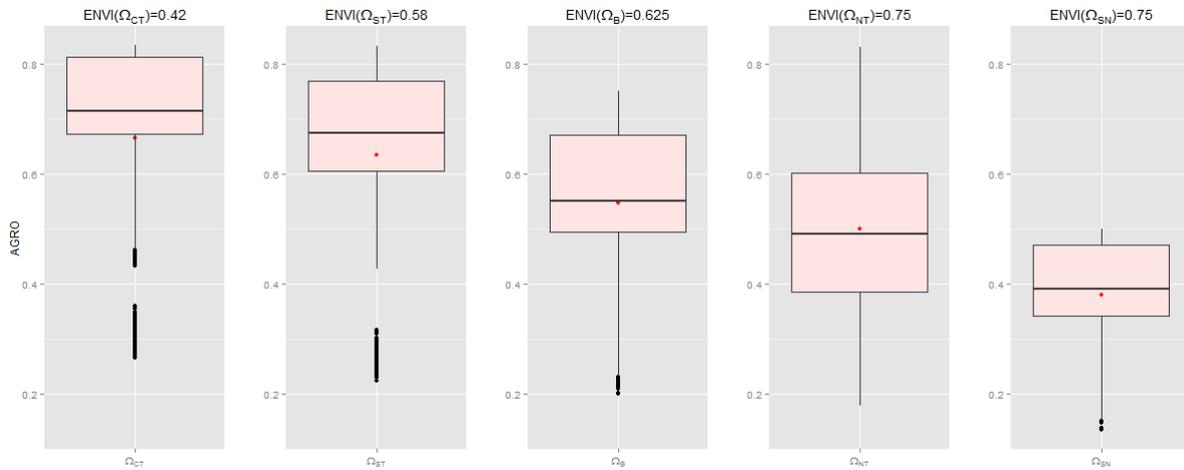

**Figure 2:** Boxplots of the agronomical performance criterion *AGRO* as a function of the five land-use distributions $\boldsymbol{\Omega_B}$ (balanced), $\boldsymbol{\Omega_{SN}}$ (half occupied by semi-natural habitats), $\boldsymbol{\Omega_{NT}}$ (half occupied by not treated crops), $\boldsymbol{\Omega_{ST}}$ (half occupied by slightly treated crops) or $\boldsymbol{\Omega_{CT}}$ (half occupied by conventionally treated crops). The red points correspond to the mean *AGRO* value of each land-use distribution, the value of the environmental criterion of each land-use distribution is given for reference above each boxplot. in each boxplot, the red point indicates the mean.

## IV. 2 Effect of crop rotation on landscape productivity

Rotation frequency ($\mu$) has significant effects on agronomical performances only when pest populations has low growth rate or/and weak dispersion ability. When pest growth rate is low ($r_P = 2\ln(2)$) there is a positive relationship between $\mu$ and AGRO (Figure 3); this relationship is even stronger when the pest diffusion coefficient is also low ($r_P = 2\ln(2)$ and $d_P = \frac{0.1}{n^2}$). This relationship can be observed for all 5 landscape compositions.

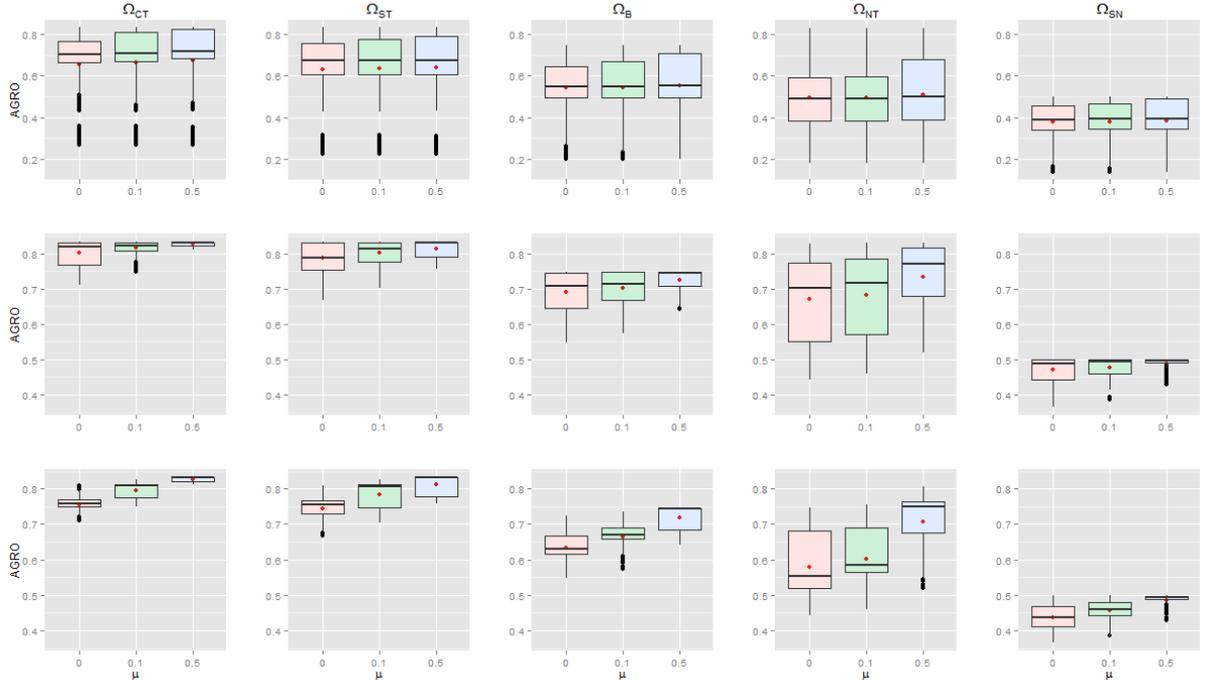

**Figure 3:** Boxplots of the agronomical performance criterion ($AGRO$) as a function of rotation index $\mu$ corresponding to the average proportion of crops (land-use types CT, ST or NT) modified in the landscape per year. The first row line corresponds to results of all simulations pooled for a given land-use distribution, the second row corresponds to the subset of simulations with low pest growth rate ($r_P = 2\ln(2)$) and the third row corresponds to the subset of simulations with both low pest growth rate ($r_P = 2\ln(2)$) and low pest diffusion coefficient ($d_P = \frac{0.1}{n^2}$). Red points indicate the mean.

## IV. 3 Relative effects of the input variables on landscape productivity

A linear model (with centered and reduced data) was computed to explain the agronomical performance as function of the model parameters:
$$AGRO \sim fr + \mu + \nu + r_P + d_P + d_N + \alpha.$$
All the statistical analyzes were performed with R software. All tested model parameters have a significant linear effect on the agronomical performance (Table 3), but the estimates of the regression parameters indicate that some parameters have a strong impact on the agronomical performance (pesticide efficiency ($\nu$), pest growth rate ($r_P$) and predation index ($\alpha$)), while the rest have a much smaller effects (rotation frequency index ($\mu$), fragmentation index ($fr$), pest diffusion coefficient ($d_P$) and natural enemy diffusion coefficient ($d_N$)). Unsurprisingly, pesticide effect and predation index have a positive effect on the agronomical criterion while pest growth rate has a negative effect. There is an interaction between the effects of pesticide efficiency and predation index: the effect of the predation rate is stronger when the pesticide efficiency is low (Table 4).

|        | Estimate | Std. Error | t value | Pr(>\|t\|)   |
|--------|----------|------------|---------|--------------|
| $fr$   | 0.02     | 1.140e-03  | 19.29   | <2e-16 ***   |
| $\mu$  | 0.03     | 1.140e-03  | 22.32   | <2e-16 ***   |
| $\nu$  | 0.44     | 1.140e-03  | 394.06  | <2e-16 ***   |
| $r_P$  | -0.56    | 1.140e-03  | -495.05 | <2e-16 ***   |
| $d_P$  | 0.05     | 1.140e-03  | 44.81   | <2e-16 ***   |
| $d_N$  | -0.02    | 1.140e-03  | -17.17  | <2e-16 ***   |
| $\alpha$ | 0.22   | 1.140e-03  | 196.37  | <2e-16 ***   |

**Table 3:** Linear regression of **AGRO** as a function of fragmentation index ($fr$), rotation frequency index ($\mu$), pesticide effects ($\nu$), pest growth rate ($r_P$), pest diffusion coefficient ($d_P$), natural enemy diffusion coefficient ($d_N$) and predation index ($\alpha$).

|              | $\nu = 2\ln(2)$ | $\nu = 2\ln(50)$ | $\nu = 2\ln(100)$ |
|--------------|-----------------|------------------|-------------------|
| $\alpha = 0$   | 0.35±0.22       | 0.58±0.17        | 0.59± 0.17        |
| $\alpha = 8/3$ | 0.39±0.21       | 0.59±0.16        | 0.60± 0.16        |
| $\alpha = 8$   | 0.53±0.15       | 0.64±0.13        | 0.65±0.13         |

**Table 4:** Mean values of the **AGRO** criterion (±sd) as function of different combination of predation index $\alpha$ and pesticide index $\nu$ values.

For the parameters with small effects, the interpretation of the effects is less straightforward: both fragmentation and rotation frequency increase the agronomical criterion. The diffusion coefficient of the pest also increases the agronomical criterion while the natural enemy's diffusion coefficient reduces it. There is an interaction between fragmentation and predation index: an increased fragmentation increases slightly the agronomical performance for all configurations, but only if predation can occur (Table 5).

|              | $fr = 0.1$   | $fr = 0.5$   | $fr = 0.9$   |
| ------------ | ------------ | ------------ | ------------ |
| $\alpha = 0$   | 0.50±0.22    | 0.50± 0.21   | 0.50± 0.21   |
| $\alpha = 8/3$ | 0.52± 0.21   | 0.53± 0.20   | 0.53± 0.20   |
| $\alpha = 8$   | 0.60± 0.15   | 0.61± 0.14   | 0.61±0.14    |

**Table 5:** Values of mean value of the ***AGRO*** (±sd) criterion as a function of the predation index $\alpha$ and the fragmentation level $fr$.